\documentclass[epj,nopacs]{svjour}
\usepackage{amsmath,latexsym,amssymb}
\usepackage{hyperref}
\usepackage{graphicx}
\newcommand{\be}{\begin{equation}}
\newcommand{\ee}{\end{equation}}
\newcommand{\bea}{\begin{eqnarray}}
\newcommand{\ena}{\end{eqnarray}}

\newcommand{\ben}{\begin{displaymath}}
\newcommand{\een}{\end{displaymath}}
\newcommand{\ba}{\begin{eqnarray}}
\newcommand{\ea}{\end{eqnarray}}
\newcommand{\ban}{\begin{eqnarray*}}
\newcommand{\ean}{\end{eqnarray*}}
\newcommand{\bs}{\begin{split}}
\newcommand{\es}{\end{split}}

\begin{document}

\title{Mirror Matter, Mirror Gravity and Galactic Rotational Curves}

\author{Zurab Berezhiani\inst{1,2} \and Luigi Pilo\inst{1,2} \and Nicola
    Rossi\inst{1,2} } 

\institute{Dipartimento di Fisica, Universit\`a  di L'Aquila 67010 Coppito, AQ, Italy \and
  INFN, Laboratori Nazionali del Gran Sasso, 67010 Assergi, AQ,  Italy} 

\abstract{
We discuss astrophysical implications of  the modified gravity model 
in which the two matter components,  ordinary and dark,  
couple to separate gravitational fields 
that mix  to each other through small  mass terms.  
There are two spin-2 eigenstates:  the massless graviton that induces 
universal Newtonian attraction,  and the massive one  that gives rise to the 
Yukawa-like  potential which is repulsive between the ordinary and dark bodies. 
As a result the distances much smaller than the Yukawa radius $r_m$ 
the gravitation strength between the two types of matter becomes vanishing. 
If $r_m \sim 10$ kpc,  a typical size of a galaxy, 
there are interesting implications for the nature of dark  matter. 
In particular, one can avoid the problem of the cusp  that is typical for the 
cold dark matter halos.  
Interestingly,  the flat shape of the rotational curves can be explained 
even  in the case  of the collisional and dissipative dark matter (as e.g. mirror matter) 
that  cannot give the extended halos but  instead must form galactic discs 
similarly to the visible matter.  
The observed  rotational curves for the  large, medium-size and dwarf galaxies
can be nicely reproduced.  We also briefly discuss possible implications for 
the direct search of dark matter. 
}

\maketitle
 
\section{Introduction}

Observational data on the Cosmic Microwave  Background  and 
on the cosmological  pattern of matter distribution 
strongly support the presence of dark components 
in  the Universe. The cosmological energy density  being very close to the critical: 
$\Omega_{tot}  = \Omega_M + \Omega_{\Lambda} \simeq 1$,  
is dominated by  dark energy: $\Omega_{\Lambda}\approx 0.75 $,  
and  the rest is non-relativistic matter:  $\Omega_M \approx 0.25 $.  
In the latter baryons take a modest part:  $\Omega_B\approx 0.04 $, 
while the main fraction  $ \Omega_{D}=\Omega_M - \Omega_B\approx 0.21 $ 
is  attributed to dark matter, a hypothetical non-baryonic  component 
which is dark in terms of the photons but 
it  interacts with the visible matter gravitationally.    
  
Strong evidence for dark matter comes from  the galactic rotational curves. 
The  luminous mass  is mostly concentrated in the inner core (bulge) of the galaxy. 
Therefore, in the absence of dark matter,  the gravitational potential  
at large distances from the galactic center is  approximately $\phi(r) \propto  1/r$, 
and the velocity of rotating matter  is expected to fall as  $ v(r) \propto 1/ r^{1/2} $ 
(known as Keplerian fall-off). 
However, the observed rotational curves are very far from being Keplerian; 
in fact,  in most of  the disc galaxies they are nearly flat outside their bulges,  
i.e. velocities $v(r)$  become approximately constant at large $r$.  

The anomalous behavior of the rotational curves can be explained 
provided that dark matter  forms extended, approximately spherically 
symmetric  halos  with  specific density profiles \cite{bahcall,begeman,burkert}. 
N-body simulations  show that the cold dark matter (CDM), 
being  collisionless and non-dissi\-pative,  
forms extended galactic halos, but  with the density profile  
having a central singularity  (cusp),
$\rho(r) \propto r^{-a}$, where $a$ ranges from $a=1$   \cite{nfw01},
up to $a=1.5$ \cite{moore01}. 
If such cusps really exist, then one can hardly reproduce 
the observed rotational curves \cite{sal2}. 
For most of  the galaxies, the singular profiles 
should  manifest by  a steep growth of the velocity at small distances 
from the center,  which however does not agree with observations.
Clear examples are  given by the dwarf and low surface brightness (LSB) galaxies 
that should be dark matter dominated, but their rotational curves  
show no sharp rise  \cite{spekkens}.  
In this respect, the non-singular profiles \cite{begeman,burkert} 
would be more satisfactory.
The question whether the CDM indeed forms cusps or 
these are artifacts  of  the numerical simulations, is still debated in the literature. 
In particular,  one might hope that the correct account for the baryon dynamics 
 can smooth-out the singularity in dark matter profiles. 
But the problem potentially remains until complete and reliable computations 
are performed.  

Many  candidates have been proposed for dark matter: 
axion ($m\sim 10^{-5}$ eV), sterile neutrino ($m\sim 1$ keV), 
wimp ($m\sim 1$ TeV),  wimpzilla ($m\sim 10^{14}$ GeV), etc.  
but its true nature is still unknown. 
In the context of  the above candidates and  
popular scenarios for  primordial  baryogenesis 
there is no natural explanation for 
the fact that  the baryon and dark  matter fractions  are so close, 
$\Omega_{D}/\Omega_B\sim 5$. 

One plausible picture is that dark matter emerges from a hidden gauge sector (or sectors) 
which has as complex  microphysics  as the observable world itself. 
In other words, the hidden sector can have a gauge structure resembling 
the Standard Model of the strong, weak and electromagnetic interactions 
or some of its extensions. 
In addition, it can also have accidental conservation laws 
that may render some of its particles stable or very long lived, 
(with a lifetime bigger than the  age of the Universe)        
exactly like the baryon number conservation in the Standard Model 
guarantees  the proton stability.  

Popular idea of {\it mirror world}  \cite{mirror,BK,FLV}  (see also \cite{Lee-Yang,NS}) 
in fact  suggests that  the hidden sector is just a {\it duplicate} of the ordinary one  
with  exactly the same particle physics 
(for the reviews, see e.g.  \cite{alice,fractions}).
If the hidden mirror sector exists, then the Universe 
along with the ordinary particles  (photon, electron, nucleons,  etc.) 
should contain also their twins  
(mirror photon, mirror electron, mirror nucleons,  etc.).\footnote{
In refs.  \cite{mirror} -  \cite{Lee-Yang}  mirror sector was introduced 
for achieving formal parity restoration in weak interactions 
and  it was   identical to the ordinary one modulo its chirality: 
while our sector is left-handed, another sector was assumed right-handed 
-- therefore the name {\it mirror}  \cite{mirror}. 
For our discussions the concerns of parity are irrelevant: 
parallel sector can be left-handed as well, and its particles can be called 
more appropriately  as the {\it twin} electrons, the {\it twin} nucleons, etc.   
Nevertheless, in the following we shall continue to coin  them as `mirror'.  
}
Turning gravity on,  in  the context of General Relativity (GR)  
the two sectors are coupled to the same metric tensor $g_{\mu\nu}$ 
and the whole theory is given by the  Einstein-Hilbert action 
\begin{equation}
S = \int d^4 x  \sqrt{g}  \left(
\frac{M^2_{\rm P}}{2} R  + {\cal L}_1 + {\cal L}_2 
+ {\cal L}_{\rm mix}   \right) , 
\label{grav}
\end{equation}
where $M_{\rm P}$ is the reduced Planck mass and  $g=-{\rm Det} [g_{\mu\nu}]$. 
Here ${\cal L}_1(\psi_1) $ and ${\cal L}_2(\psi_2) $ are the Lagrangians respectively for  
the ordinary and mirror particles/fields $\psi_1$ and $\psi_2$,  
while the ``mixed" Lagrangian ${\cal L}_{\rm mix}(\psi_1,\psi_2)$ 
contains  possible interaction terms 
as are e.g. the  photon - mirror  photon  kinetic mixing term 
$\epsilon F^{\mu\nu}_1 F_{2\mu\nu}$ \cite{photons},   
the couplings  that induce mass mixings between the ordinary and mirror 
neutrinos \cite{neutrinos} and  neutrons \cite{neutrons,Berezhiani:2008bc},   
or  possible common gauge interactions between  the two sectors \cite{gauge}.  
Mirror parity, a discrete  symmetry under the exchange 
of the two particle sets ($\psi_1 \leftrightarrow  \psi_2$),  
guarantees that the  Lagrangians ${\cal L}_1$ and ${\cal L}_2 $ are identical, 
i.e.  ordinary (type 1) particles and mirror (type 2) particles have exactly 
the same characteristics (masses, coupling constants, etc.). 

In spite of having the same microphysics, 
the two sectors cannot have the same cosmological realization. 
In fact, the Big Bang Nucleosynthesis bounds
require that the temperature of mirror sector $T'$ must be at least 
twice smaller than that  of the ordinary sector  $T$ \cite{cosmology,BCV}. 
In addition, when $T' < T$  mirror matter can be 
a viable candidate for dark matter: 
namely, for  $T'/T \lesssim 0.2-0.3$  mirror dark matter (MDM)  
produces the same pattern for the large scale power spectrum and 
the CMB anisotropies as the standard CDM\cite{BCV,Ignatiev,BCCV}. 

In addition, if $T' < T$ in the Early Universe, the baryon asymmetry  can be generated, 
in both sectors,  
via out-of-equilibrium $B-L$ and $CP$ violating collision processes between 
ordinary and mirror particles \cite{bariogenesi}  induced by those interaction terms 
in $ {\cal L}_{\rm mix}$  that also induce the neutrino or neutron mixings 
with their mirror twins. 
Such a baryogenesis  mechanism can nicely explain the puzzling  
`coincidence'  between the visible and dark matter fractions  in the Universe,  
 predicting the ratio $\Omega_{D}/\Omega_B\sim 1 \div 10$  \cite{fractions}. 

In contrast to the CDM, mirror baryons 
constitute collisional and dissipative dark matter. 
So one expects that the MDM should undergo 
a dissipative collapse and clump in galaxies 
instead of producing extended quasi-spherical halos.\footnote{ 
If the mirror matter fragmentation and mirror star formation process 
is fast enough, the halos can be represented by the mirror elliptical galaxies:  
see discussions in \cite{BCV,BCCP}.  } 
Then, for a dark component $M_2$  being as compact as the luminous one $M_1$, 
the gravitational potential at large distances from  the galactic center  would be 
\begin{equation} \label{pot}
  \phi(r) = -  \frac{ G_{\rm N} (M_1+M_2) }{r} , \quad \quad G_{\rm N}=\frac{1}{8\pi M_{\rm P}^2}, 
\end{equation}
with $G_{\rm N}$ being the Newton constant. 
Hence, the rotational velocities should show the Keplerian fall-off  
$v(r) \propto \sqrt{1/ r} $ in the outer regions of the galaxies
together with  a steeper rise in the inner regions,  
both aspects contradicting the observations. 
Therefore,   without modifying the  gravity, 
the MDM scenario can get into serious difficulties 
in explaining the galactic rotational curves.


In a previous paper  \cite{trigravity} we suggested a model for a large distance 
Yukawa-type modification of gravity that involves two gravitational fields. 
It was assumed  that ordinary (type 1) matter and and dark mirror (type 2) matter 
interact with two separate metric tensors  $g_{1\mu\nu}$ and $g_{2\mu\nu}$, 
i.e. each sector has its own GR-like gravity.  
The effective action of this model contains 
Einstein-Hilbert terms per each sector 
and a mixing term  between the two sectors: 
\begin{eqnarray}
S = & \int d^4 x  \left[\sqrt{g_1}  \left(\frac{M^2_{\rm P}}{2} R_1  + {\cal L}_1  \right)+ 
\sqrt{g_2}  \left(\frac{M^2_{\rm P}}{2}  R_2 + {\cal L}_2  \right) 
 \right.
 \nonumber \\
& + \left.  \left(g_1 g_2 \right)^{1/4} 
\left( \varepsilon^4 V_{\rm mix}   + \mathcal{L}_{\rm mix} \right) \right] ,
\label{bg}
\end{eqnarray}
where $\varepsilon$ is a small mass scale and 
$V_{\rm mix}(X)$ is a scalar function containing  non-derivative terms 
built out of the two metrics through the unique available tensor  combination
$X^\mu_\nu = g_2^{\mu\sigma} g_{1\sigma\nu}$. 
In particular,  one can have in $V_{\rm mix}$ 
terms  like ${\rm Tr}(X+X^{-1})$,  ${\rm Tr}(X^2+X^{-2})$, etc.    
(cosmological terms can be also included as 
$ \Lambda \, ({\rm det }X+ {\rm det }X^{-1})$). As for  
$\mathcal{L}_{\rm mix}(\psi_1,\psi_2,X) $, it  again describes possible 
interaction terms between the ordinary and mirror particles.  
Now `mirror' parity along with  the two sets of matter fields ($\psi_1 \leftrightarrow  \psi_2$) 
exchanges also the two metric fields ($g_{1\mu\nu} \leftrightarrow g_{2\mu\nu}$, i.e. 
$X \leftrightarrow X^{-1}$).   

The theory admits a Lorentz-invariant vacuum solution 
$\eta_{\mu\nu} = {\rm diag}(-1,1,1,1)$ for both metrics.\footnote{
For having a flat solution, the effective vacuum energy has to be 
fined tuned, as  in the case of the GR.
}   
As a result, the two gravitational fields  
$h_{1\mu\nu}=g_{1\mu\nu}-\eta_{\mu\nu}$ and 
$h_{2\mu\nu}=g_{2\mu\nu}-\eta_{\mu\nu}$ 
 couple  to  the respective   energy momentum tensors $T_{1\mu\nu}$ and $T_{2\mu\nu}$.   
The terms in $V_{\rm mix}(X)$ induce a mass mixing between the two gravitons. 
Notice that in the presence of the mixing term $V_{\rm mix}$ only  the 
{\it diagonal}  diffeomorphisms are unbroken. Hence,  one eigenstate   
$h_{\mu\nu} = \frac{1}{\sqrt2} ( h_{1\mu\nu} + h_{2\mu\nu})$ ({\it true} graviton), 
that couples to both sectors symmetrically through the combination
$T_{1\mu\nu} + T_{2\mu\nu}$,  remains massless;  a second eigenstate  
$f_{\mu\nu} =\frac{1}{\sqrt2} ( h_{1\mu\nu} - h_{2\mu\nu})$ 
which couples to the antisymmetric combination $T_{1\mu\nu} -
T_{2\mu\nu}$ gets a mass  $m_f \sim \varepsilon^2/M_{\rm P}$.

The theoretical consistency requires that the massive spin-2 
field must have Lorentz-violating mass pattern. For this, in ref. \cite{trigravity}  
a third  metric tensor $g^3_{\mu\nu} $ was introduced  
that couples to both metrics  in (\ref{bg}). So,   the potential 
$V_{\rm mix}$, apart of the combination 
$X^\mu_\nu = g_2^{\mu\sigma} g_{1\sigma\nu}$, 
also includes the combinations $g_1^{\mu\sigma} g^3_{\sigma\nu}$
and $g_2^{\mu\sigma} g^3_{\sigma\nu}$  
that induce a Lorentz-violating mass terms for the eigenstate  $f_{\mu\nu} $ when
 the third metric gets the Lorentz-breaking vacuum 
configuration  $\eta^3_{\mu\nu} = {\rm diag}(-c^2,1,1,1)$ \cite{pilo}.  
(Technically, the third metric can be rendered non dynamical in a
suitable limit, and the dynamical effects of the `third' graviton can be neglected). 
The same goal  can be achieved 
by a dynamical Lorentz-breaking condensate \cite{Berezhiani:2008ue}. 
The models of massive gravity (bigravity) with the Lorentz invariance \cite{Isham,damour}
have serious  theoretical problems while giving up the Lorentz invariance allows to 
 have a healthy ghost-free theory.\footnote{
Under certain assumptions, one can prove a  theorem  \cite{Boulanger:2000rq} 
that there cannot exist consistent (ghost-free) interactions of two massless  gravitons. 
This proof, however, does not apply to our model \cite{trigravity} where the spin-2 field 
$f_{\mu\nu}$ is massive and moreover its mass terms are Lorentz non-invariant.
}
In addition, such a theory does not suffer from 
the Van Dam-Veltman-Zakharov discontinuity problem 
\cite{trigravity,pilo,rubakov}. 
Regarding the experiments for  the light deflection and light travel time 
our model has no deviation from the GR predictions 
(i.e. in terms of the Post-Newtonian parameters $\gamma_{\rm PPN}=1$). 
The exact spherically symmetric solutions for a Lorentz-breaking 
massive bigravity were found in ref. \cite{pilo2}.  
The parametrized Post-Newtonian limit of the bigravity theories were 
recently studied in ref. \cite{Clifton}. 

Theoretical aspects  of the model are discussed in details  in \cite{trigravity,nic}. 
For what concerns us,  the bottom line is that  it  leads to interesting 
long-distance modification of the gravity. 
Namely,  taking a point-like source composed of both type 1 and type 2 masses,  
$M_1$ and $M_2$,  the exchange of massless graviton 
$h_{\mu\nu}$ induces the following potential  for the both type 1 and type 2 test particles:
\begin{equation} \label{S}
  \phi_h(r)= - 
  \frac{G_{h}(M_1 + M_2 )}{r}\,  , \quad\quad 
  G_h =  \frac{1}{16\pi M_{\rm P}^2}
  \end{equation}
On the other hand, the massive graviton  $f_{\mu\nu}$ coupled to  
the anti-symmetric combination of $M_1$ and $M_2$  induces the following 
Yukawa-like potential for a type 1 test particle:  
\begin{equation} \label{bigr}
  \phi_f (r)= - 
  \frac{G_{f}(M_1 - M_2 )   e^{-\frac{r}{r_m}} }{r} \,  , 
  \quad\quad   G_f =  \frac{1}{16\pi M_{\rm P}^2}  
 \end{equation}
while for a type 2 test particle the potential is $-\phi_f(r)$. 
The Yukawa radius is defined by the Compton wavelength of the 
massive graviton, $r_m = 1/m_f$. 
This potential is {\it repulsive} between the type 1 and 
type 2 matter.  
As far as $G_f=G_h=G_{\rm N}/2$,  the  universal attraction (\ref{S}) between the latter
 is exactly cancelled by this repulsion in the limit $r/r_m \to 0 $.  
Hence,  at small distances, $r\ll r_m$,  there is practically  no gravitation 
between the ordinary matter and dark matter  objects.\footnote{
Let us remark that the equality $G_f=G_h$  is not generic: 
 e.g., for  $f_{\mu\nu}$ with the Lorentz-invariant (Fierz-Pauli) mass term  
 one would have $G_{f} =\frac43 G_{h}$ and a sick theory suffering 
 from the discontinuity problem and other diseases.
 In our model $G_f=G_h$  is due to the Lorentz-breaking  in the mass terms 
 for  $f_{\mu\nu}$.  
In general  case the Lorentz-breaking originates {\it two} Yukawa terms  of the type (\ref{bigr})
 with the different radii  $r_{1m}$ and $r_{2m}$  \cite{trigravity}. 
 In this paper, for the sake of simplicity,  we take  them equal, $r_{1m}=r_{2m}=r_m$.
 }

The full potential felt by an ordinary  test particle 
$\phi(r) = \phi_h(r) + \phi_f(r)$ can be presented as:   
\begin{equation} 
\label{bigr2}
  \phi(r)=  - \frac{G_{\rm N} (1 +  e^{-\frac{r}{r_m}})M_1}{2r}  
  - \frac{G_{\rm N} (1 - e^{-\frac{r}{r_m}})M_2 }{2r} \, , 
\end{equation}
while the corresponding potential for a type 2  test particle, $\phi_-(r) = \phi_h(r) - \phi_f(r)$, 
is obtained from (\ref{bigr2}) by exchange $M_1 \leftrightarrow M_2$.  
Hence, at small distances   the universal character  of gravity 
between the  ordinary and dark components is violated.  
This is acceptable  as far as there are 
no experimental  data regarding the gravitational interactions 
of dark matter at small distances.  
On the other hand,   at  $r \ll r_m$  gravity between type 1 bodies  
(as well as between type 2 bodies)  remains essentially standard:  
one has the  Newton law  with a constant $G_{\rm N}$   
and the weak equivalence principle is respected for all kinds of visible matter. 
These properties are verified with a spectacular precision in a wide range of 
distances:  from a fraction of mm (terrestrial experiments)  
 to several AU (solar system tests). Therefore, 
 our potential (\ref{bigr})  does not  conflict with the experimental data 
 provided that the Yukawa radius is much larger than the solar system,  
say  $r_m \gg 20~ {\rm AU} = 10^{-4}$ pc. 

At large distances,  $r \gg r_m$, the massive spin-2 field decouples:  
the Yukawa term $\phi_f(r)$ (\ref{bigr}) dies out and 
only the long-range potential $\phi_h(r)$ (\ref{S})  remains at work, which is 
universal between type 1 and type 2 components. 
Obviously, the latter is just the same as the familiar Newtonian potential 
 (\ref{pot})  but with a {\it halved} Newton constant $G_h= G_{\rm N}/2$.  
Variation of the Newtonian constant with the distance, however,  would 
 not contradict the observational data if the Yukawa radius is larger than few pc, 
 the distance scale at which the systems of the gravitationally bounded stars 
 still can be tested.  
 
Interesting features  can emerge  at intermediate distances 
$r \sim r_m$  where both potentials (\ref{S}) and  (\ref{bigr}) are important. 
In particular, in this paper we take $r_m \sim 10$ kpc, 
a typical optical size of galaxies.  This will allow us to explain the  
nearly flat shape of the galactic rotation curves  even if dark matter 
does not form extended halos but it collapses into discs similarly 
to the visible component, exactly what is expected in the case of mirror matter.   
 
 The paper is organized as follows. In the next section we 
discuss the case of  point-like type 1  and type 2 sources and 
the corresponding modification of the Newton law at large distances. 
In section 3  we study the implications for the galactic rotational curves. 
Finally, in section 4, we shortly discuss our results and outline the implications 
of our scenario for the direct search of dark matter.

\section{Modified Newton Law}

In this section we show that  at distances  $r\sim r_m$, 
the typical rotational velocities of an ordinary test body in the
gravitational field produced by mirror matter has a rather flat shape 
even if the source is  point-like. 

\begin{figure}
\includegraphics[width=.48\textwidth]{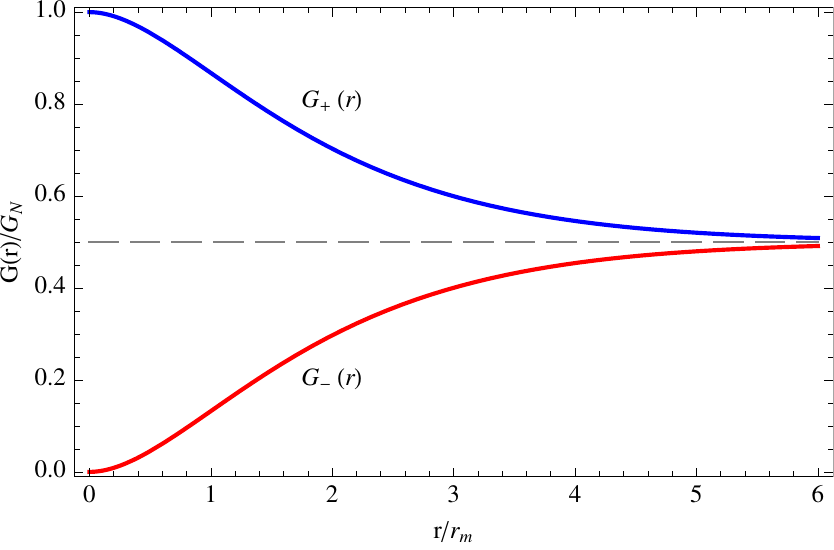}
\caption{
The distance dependent "constants" $G_+(r)$  (upper curve) and $G_-(r)$ (lower curve),  
in units of the experimental Newton constant $G_{\rm N}$.}
\label{Newt-mir}
\end{figure}

The free fall acceleration of a type 1  test  particle in the  potential (\ref{bigr2}) 
reads:\footnote{
Observe that for $M_1=M_2$ the Yukawa term becomes ineffective and 
the acceleration is exactly the same as the one induced by the source $M_1$ 
in the case of normal Newtonian gravity: $g(r)= G_{\rm N} M_1/r^2$.
}  
\begin{eqnarray} 
\label{g}
 g(r) = && 
  \frac{G_{\rm N} }{2}  \left[ \frac{M_1+M_2}{r^2} +\frac{M_1-M_2}{r^2}
  \left(1+\frac{r}{r_m} \right) e^{-\frac{r}{r_m}} \right]  
  \nonumber \\
  && =  \frac{G_+(r)M_1}{r^2} +  \frac{G_-(r)M_2}{r^2}   
 \end{eqnarray}
where  the effective ``Newton" constants $G_\pm$  become  distance dependent
functions: 
\begin{equation}
\frac{G_\pm (r)}{G_{\rm N}} = \frac{1}{2}\left[ 1 \pm  \left(1+\frac{r}{r_m}\right) e^{-\frac{r}{r_m}}\right] , 
\label{Gpm}
\end{equation}
so that $G_+(r) + G_-(r) = G_{\rm N}$ (see Fig. \ref{Newt-mir}). Namely,  
$G_{ +}(r)$  measures  the gravitational strength between the two type 1 objects 
(and between two type 2 objects) while  $G_{-}(r)$ the strength between 
the type 1 and type 2 objects.   

At small distances, $r \ll r_m$,  we have: 
\begin{equation}
\frac{G_+ (r)}{G_{\rm N}}  \approx  1 -  \frac14 \left( \frac{r}{r_m}\right)^2 , 
\quad\quad 
\frac{G_-(r)}{G_{\rm N}}  \approx   \frac14 \left( \frac{r}{r_m}\right)^2 . 
\label{Gpm-small}
\end{equation} 
In the limit $r\to 0$ we get  $G_+=G_{\rm N}$ while $G_-$ vanishes.  
Thus, at small distances the gravitational force between  
ordinary objects  obeys the universal $1/r^2$ Newton low with a constant $G_{\rm N}$,  
while  dark matter becomes  dark also for the gravity.  
More precisely,  at  $r\ll r_m$ 
the acceleration of a test particle in the field of dark point-like source  $M_2$ 
is constant but very small,  $g_-(r) = G_{\rm N}M_2/4r_m^2$. 

At large distances, $r \gg r_m$,  we get 
\begin{equation}
\frac{G_+ (r)}{G_{\rm N}} = \frac{G_- (r)}{G_{\rm N}} = \frac12   
\label{Gpm-large}
\end{equation} 
with the precision of exponentially small terms. 
Thus, in the limit $r \to \infty$,  
the gravitational force between  ordinary and dark matter becomes universal and 
$\propto 1/r^2$ but with a {\it halved} Newton constant $G_{\rm N}/2$.  

For  $r_m \sim 10$ kpc, the gravity modification effects  are too small and  
have no impact for the solar system tests: 
eq. (\ref{Gpm-small}) shows that the sun and its planets gravitate 
with the same Newton constant $G_{\rm N}$ that is experimentally measured 
in small distance experiments on the Earth,  
up to  correction less than  (20 AU/10 kpc)$^2 \sim 10^{-16}$.  
On the other hand, dark (mirror) objects have almost no attraction to the sun   
and within the solar system their acceleration is practically vanishing:  
$g_2 = G_{\rm N}M_\odot/4 r_m^2 \approx 4 \times 10^{-20}$ cm/s$^2$.  

On the other hand, at large distances the exponential term in (\ref{Gpm}) dies out  and
 e.g.  for $r > 1$ Mpc,  within a precision better than $10^{-41}$
 we recover the universal $1/r^2$ force law but with the {\it halved} Newton constants  
 $G_\pm = G_{\rm N}/2$.  
Thus, as far as  the present cosmological expansion 
and formation of the large scale structures are concerned  
for which the horizon crossing scales are much larger than $r_m\sim 10$ kpc, 
all should remain nearly as in the standard FRW cosmology 
with the Newton constant being $G_{\rm N}$ at all distances. 
However,  for a given value of the Hubble constant, 
the total energy density of the Universe in the context of our model should be 
$\rho_{\rm our} = 3H_0^2/4\pi G_{\rm N}$ 
instead of $\rho\approx \rho_{\rm cr} = 3H_0^2/8\pi G_{\rm N}$.  
Hence, $\rho_{\rm our}\approx 2\rho_{\rm cr}$  \cite{trigravity}. 
Moreover,  once in our scenario  the large scale structures 
with  comoving scales larger than  few Mpc   
are governed by a \emph{halved} Newton constant, $G_{\rm N}/2$, 
it is natural to expect that 
the proportion between the dark matter and baryon densities 
in the Universe is also a factor of 2 larger than the ratio $\rho_{D}/\rho_B\simeq 5$
required in the standard cosmological `concordance' model.   
In other words, 
in order to keep the gravitational dynamics of  large structures 
like  galaxy clusters and superclusters compatible with the observations,  
one must take $\beta_{\rm cosm} = \rho_2/\rho_1 \simeq 10$.     
As in the standard cosmology, in our scenario  the dominant  contribution 
in total energy density of the Universe should come from some sort of dark energy.

\begin{figure}
\includegraphics[width=.48\textwidth]{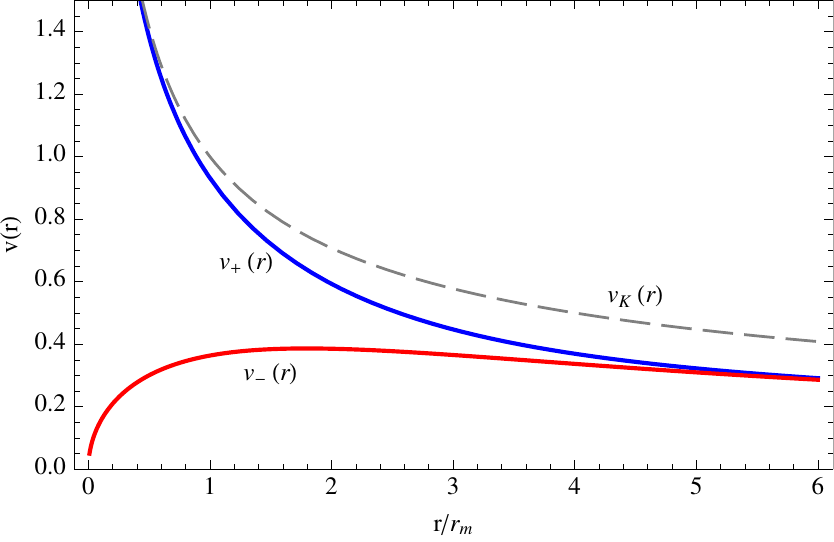}
\caption{Rotational velocities  in our model (arbitrary units) 
for  a point-like  ordinary (type 1) source with a given mass 
$M_1=M$ ($v_+$, upper solid) and for  a  mirror  (type  2)  source 
with $M_2=M$ ($v_-$, lower solid). 
    For a composite source with $M_1=M_2=M$, 
the rotational curve exactly coincides with the Keplerian one, 
$v_{\rm K}(r) = \sqrt{G_{\rm N} M / r }$, originated by the mass $M$ 
in the standard Newtonian gravity (dash).   }
                 \label{Rot-curves}
\end{figure}

However, at intermediate distances the Yukawa-like term is essential:  
for  $r\sim r_m$  the constants $G_+$ and $G_-$ are comparable but not equal. 
For $r_m \sim 10$ kpc, this can have interesting consequences for the 
testing the dark matter in the galaxies. 

In order to grasp the basic idea behind the shape of  the rotational curves 
produced by the potential (\ref{bigr2}), let us imagine a galactic object 
moving along a circular orbit of radius $r$ from the center. 
Its velocity is determined by equating the centrifugal force with the radial 
component of gravitational pull,  i.e. $v^2(r)/r = g(r)$. 
Therefore, in the Newtonian case,  $g(r)= G_{\rm N} M /r^2$ the
rotational velocity is  Keplerian: $v(r) = (G_{\rm N} M / r)^{1/2}$; 
$M=M_1+M_2$  represents the total mass of the source, regardless is
dark or visible. 

In our case, the acceleration (\ref{g}) has to be  used and  
the situation is different. For an ordinary source $M_1$ we have 
$v_+(r) = [G_{+}(r) M_1 / r]^{1/2}$, and  the velocity
fall-off  at  distances $r\sim r_m$  is even faster than in the Keplerian case 
(see upper solid curve in Fig. \ref{Rot-curves}). 
However,  the rotational velocity in the field of a mirror source $M_2$, 
$v_-(r) = [G_{-}(r) M_2 / r]^{1/2}$,  has a quite different shape. 
At small distances it increases (approximately  as $r^{1/2}$)     
approaching  a rather smooth plateau around $r\sim r_m$ and 
then decreases very slowly  (see lower curve in Fig. \ref{Rot-curves}).  

This gives a hint  that in our modified gravity model one can reproduce the
observed  rotational curves  if  $r_m \sim 10$ kpc even without requiring that 
dark matter forms extended halos. 
Instead,  it can be distributed in the galaxy in a rather compact way 
as it is  natural for mirror matter. 
Fig. \ref{Rot-curves} shows that  a typical rotational curve produced by a dark source 
is rather flat   within a wide range of distances, $r/r_m \sim 1\div 5 $,  
even if the  source is point-like. 
(Obviously, the curves should flatten more when the dark source 
is extended with a radius comparable to $r_m$.) 
In addition, Fig. \ref{Rot-curves} gives also an idea 
that  the dark matter cusp, if it exists,  can become harmless for the observations:  
it will be gravitationally `invisible' for the stars rotating closer to the galaxy 
center and the rotational curves at small distances  will not be affected. 
Needless to say, for achieving the correct rotational profiles 
 the amount of dark matter $M_2$ in the galaxy should be 
considerably bigger than that of visible matter $M_1$.

\section{Galaxy Rotational Curves}
Rotational curves describe  the velocity $v(r)$ of gravitationally bounded galactic 
objects (typically stars and interstellar gas) 
as a function  of the  distance $r$ from the center. 
The square of the rotational velocity $v^2$  has then two contributions: 
one from the visible matter and another  from the dark matter:  
\begin{equation}\label{vis-dm}
   v^2(r)  = v^2_{\rm vis}(r) +v^2_{\rm dm}(r)  \; .
\end{equation}
The form of $v^2_{\rm vis}$ depends on 
the distribution of matter in the galaxy.
In ``honest'' disk galaxies  
visible matter is distributed along the disk with the profile
(minimum disc  hypothesis\cite{freeman}):
\begin{equation}
\sigma_1(r)=\frac{M_1  }{2 \pi r^2_1} \, e^{-\frac{ r}{r_1}} .
\label{vis}
\end{equation}
The central density is normalized  by the total visible mass  $M_1$ 
and $r_1$ is the length scale of the disc. 
Thus, in the absence of dark matter,  the rotational velocities for $r\gtrsim2 r_1$  
are expected to fall off according to  Keplerian behavior  $v(r) \propto \sqrt{1/r}$. 
One observes instead that in the outer regions of  galaxies the  
rotational curves flatten, i.e. $v(r)$ is approximately constant. 
More precisely, for all spiral 
galaxies the rotational curves are not really  flat but but still far
from the expected Keplerian fall-off. 
Empirically, the observational data 
of many galaxies are  well fitted by 
the so called  universal rotational curves 
\cite{per_sal}.

\begin{figure}
\hskip -.2cm
 \includegraphics[scale=.38]{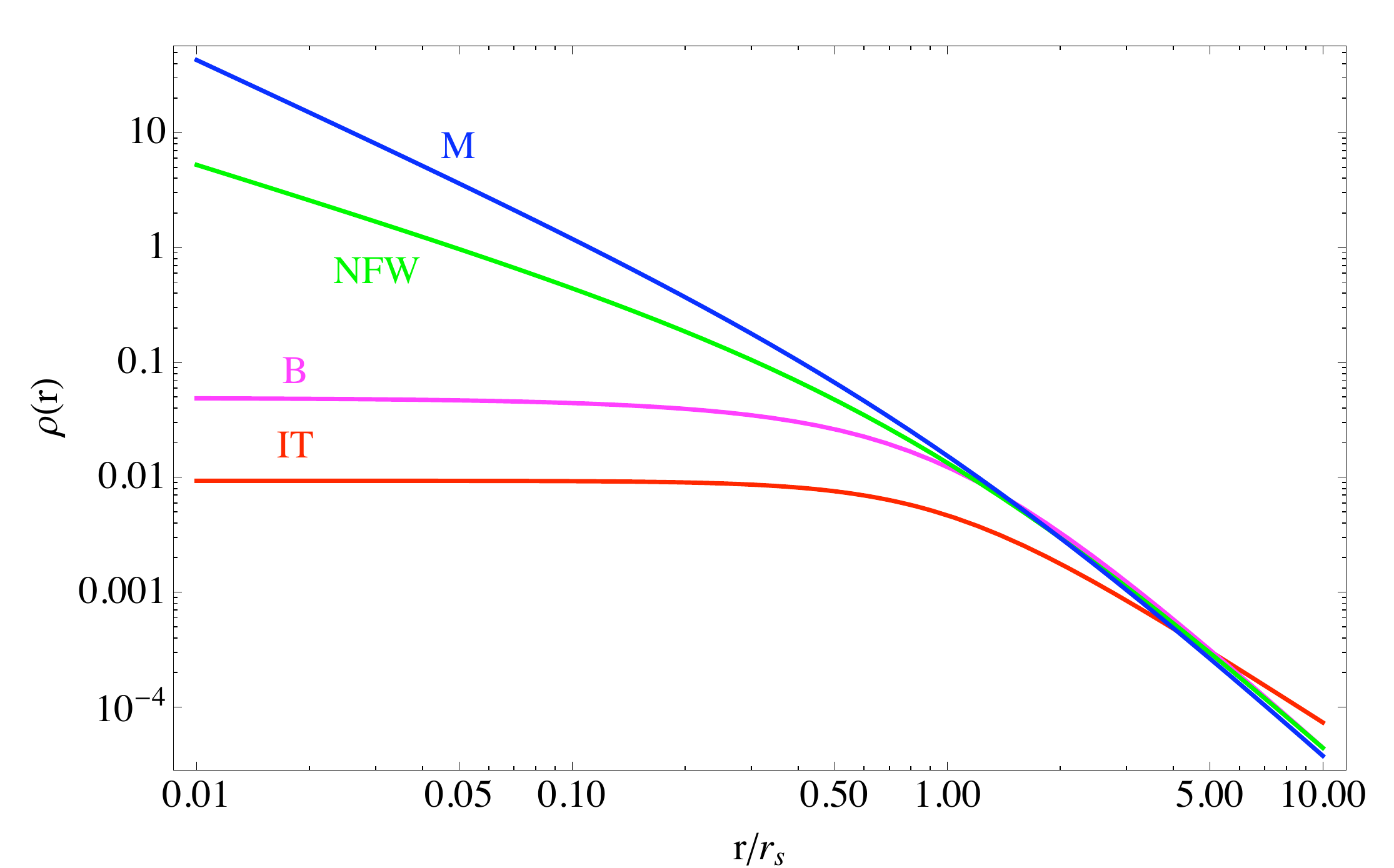}
      \includegraphics[scale=.375]{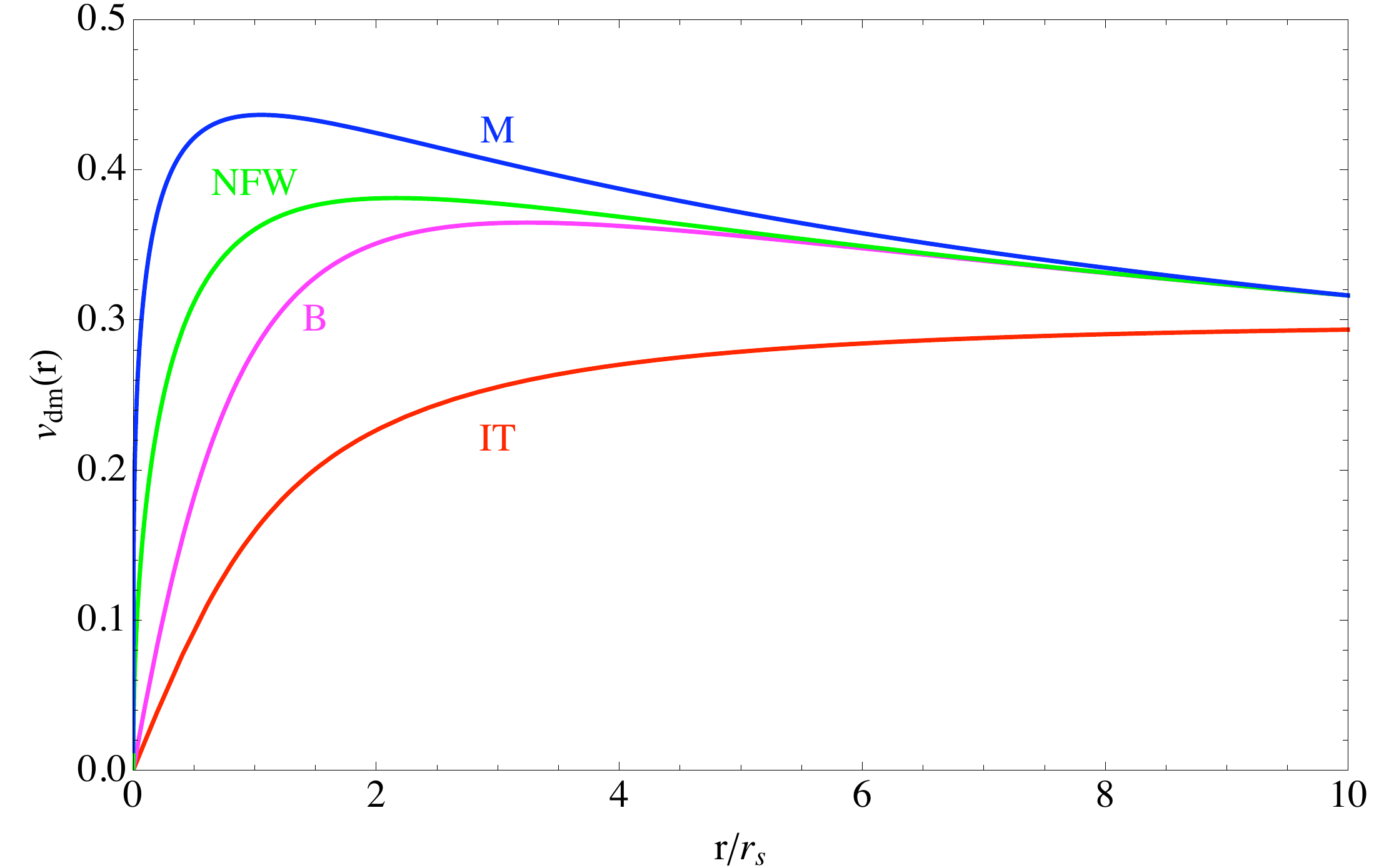}
     \caption{Density profiles $\rho(r)$  for the IT, Burkert (B), NFW and Moore (M) models
     and the respective curves for  $v_{\rm dm}(r)$.    }
   \label{nfw_2}
\end{figure}

The dark matter distribution in galaxies  depends on its nature but also 
on various subtle  details of the galaxy formation.  
Several models are considered in the literature. 
For example, the pseudo-isothermal (IT) profile \cite{begeman} 
is obtained by taking  an isothermal equation of state 
for matter in hydrostatic equilibrium in the presence of Newtonian gravity  
with a suitable boundary conditions. It has a form 
\begin{equation}\label{IT}
\rho(r) = \frac{\rho_s  }{1+ \frac{r^2}{r_s^2}} , 
\end{equation}
where $\rho_s$ is the central density and $ r_s $ is a characteristic  length scale   
defining the size of the constant density core.  
At large distances, $r \gg r_s$, the density drops as $r^{-2}$. 
Therefore, at $r\gg r_s$ the mass $M(r) = \int_0^r 4\pi r^2 \rho(r) dr$ 
grows linearly with $r$, and hence the rotational velocity 
$v_{\rm dm}(r) = \sqrt{G_{\rm N} M(r)/r}$ 
becomes constant at large distances.  For realistic cases, the 
divergence of $M(r)$ should be cut-off at some proper distance 
from the galaxy center. 

Another  core-like profile was suggested by Burkert \cite{burkert}: 
\begin{equation}
\label{B} 
	\rho(r) = \frac{\rho_s}{(1+\frac{r}{r_s})(1+\frac{r^2}{r_s^2})} . 
\end{equation}
In this case,  for large $r$ the density falls as  $\rho \propto r^{-3}$,  
and thus $v_{\rm dm} \propto (\ln r/r)^{1/2}$.  

Numerical N-body simulations for the CDM 
produce the following  density profiles: 
\begin{equation}\label{NFW}
\rho(r) =  \frac{ \rho_s}{\left(\frac{r}{r_s} \right)^{a}
  \left[1+\left(\frac{r}{r_s}\right)^a \right]^{\frac{3}{a} -1} }   , 
\end{equation}
which have a singularity (cusp) for small $r$: $\rho \sim r^{-a}$, 
with the value $a$ depending on the simulation details. 
In particular, $a=1$ for the Navarro-Frenk-White (NFW) model  \cite{nfw01}  
and $a=1.5$ for the case of Moore et al.  \cite{moore01}. 
For large $r$ one has  $\rho \propto r^{-3}$ in both cases.
%
Different  density profiles for dark matter and their 
contributions $v_{\rm dm}(r)$ to the rotational velocities 
are schematically shown in Fig. \ref{nfw_2}.

From the empirical side, the non-singular  IT or B  profiles can nicely fit  most of the 
observed  rotational curves,  even those for the dark matter  dominated  galaxies
as are the dwarf and LSB galaxies \cite{begeman,burkert}.  
However, from the theoretical side,  these core-like profiles 
are not favored by the numerical computations in the CDM picture which point instead  
towards the singular profiles (\ref{NFW}) that predict too steep rise of the rotational 
velocity at small distances which feature should be more prominent  in the case of 
the galaxies dominated by dark matter. 
This however  is in contradiction with the shallow shapes of the rotational 
curves observed for the dwarf and LSB galaxies \cite{sal1}. 

The following remark is in order. The fits show that, with the exception of the 
low-luminosity dwarfs, the visible component makes the dominant contribution 
to the rotational velocities within the bright optical disc \cite{begeman}. 
The appearance of flat rotational curves does in general require a careful 
matching of the falling contribution $v_{\rm vis}(r)$  from the disc and 
the rising contribution $v_{\rm dm}(r)$ from the halo -- the ``conspiracy". 
In the dwarf and LSB systems this conspiracy generally breaks down 
as far as the halo contribution $v_{\rm dm}(r)$ becomes dominant 
already within the optical disc.

Let us now discuss how the rotational curves can be obtained using the 
modified gravitational potential (\ref{bigr2}),  without assuming  extended  dark halos. 
The key proposal is to use  dark matter the  {\it similar} density profiles 
 for luminous and dark matter. 
Hence,  the two components are assumed to be distributed 
in two overlapping concentric disks with similar density profiles.  
Namely, we take exponential distribution (\ref{vis}) for visible matter    
and assume the analogous form  for dark matter obtained by rescaling 
 $M_1 \to M_2$ and $r_1 \to r_2$: 
\be
\sigma_1(r)=\frac{M_1  }{2 \pi r^2_1} \, e^{-\frac{ r}{r_1}} , \quad \quad 
 \sigma_2(r)=\frac{M_2  }{2 \pi r^2_2} \, e^{-\frac{ r}{r_2}} ,
\label{distr}
\ee
where  $M_{2}$ is a total dark mass in the galaxy and 
$r_2$ is  length scale of its distribution.   
For mirror dark matter this hypothesis is quite natural: 
it has the same microphysics as ordinary (baryonic) matter; 
it is then conceivable that the very same mechanism of
galaxy formation operates in both sectors giving the same
matter distributions up to a rescaling of the relevant parameters. 

\begin{figure}
\includegraphics[scale=0.75]{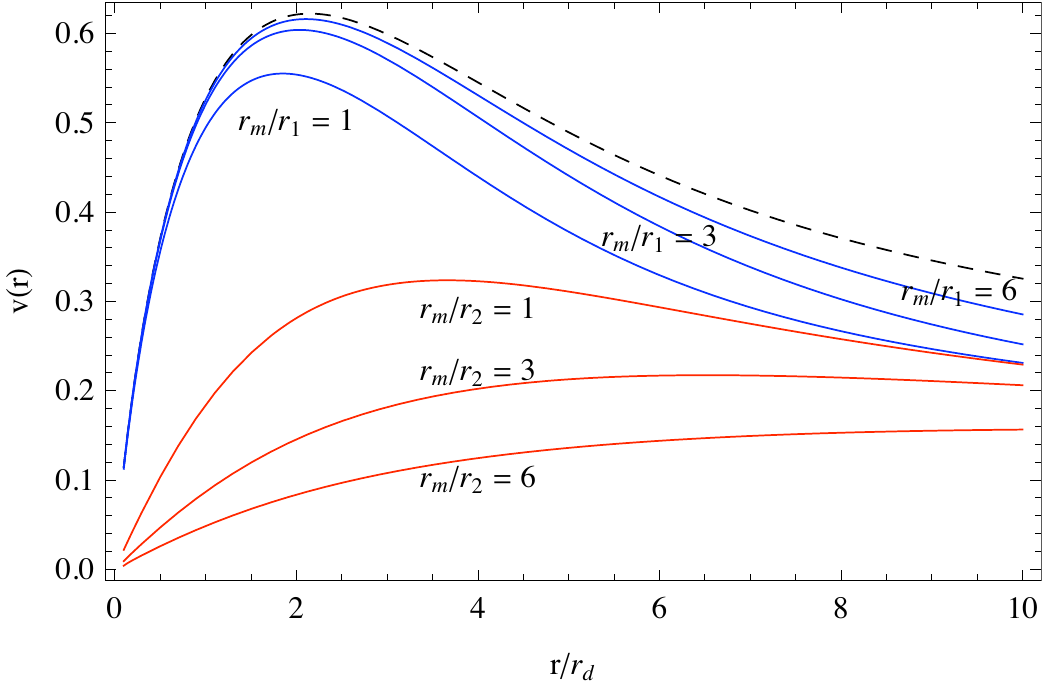}
\caption{Contributions to rotational velocities in our model 
(arbitrary units).  The dash curve shows the rotational velocity 
in the visible matter disc with a radius $r_d=r_1$ in the case of normal Newtonian gravity. 
The three upper solid curves correspond to $v_{\rm vis}$ for the same ordinary disc 
in our model  for different values of $r_m$, and the three lower solid curves 
correspond to $v_{\rm dm}$ for the mirror matter disc with $r_d = r_2$. 
  }
                 \label{Comparison}
\end{figure}

The acceleration of a test type 1 particle located at the coordinates ${\bf r}=(x,y)$ 
on the disc can be readily found:  
\begin{eqnarray} 
  {\vec g}({\vec r})=    \frac{G_{\rm N}}{2}   \int_{_{\rm Disk}} \!\!\!
\!  \! \!  r'dr'd\vartheta  
\left.  \frac{({\vec r}-{\vec r'})}{|{\vec r}-{\vec r}'|^3}
  \right[ \sigma_1({\vec r}')+\sigma_2({\vec r}') 
  \nonumber \\
+ \left.  \big[ \sigma_1({\vec r}')-   \sigma_2({\vec r}') \big]
\left(1+\frac{|{\vec r}-{\vec r}'|}{r_m}\right)
    e^{-\frac{|{\vec r}-{\vec r}'|}{r_m}} \right] , 
\end{eqnarray}
where $|{\vec r-\vec r'}| = \sqrt{r^2+r'^2-2 r r'\cos \vartheta}$.
Due to cylindrical symmetry, $v(r) = \sqrt{r g(r)} $  will depend only on 
the radius $r=(x^2+y^2)^{1/2}$.  
For the exponential density distributions (\ref{distr}) the integration can be 
performed  semi-analytically. 

In Fig. \ref{Comparison} we show typical shapes for the ordinary and 
dark contributions, $v_{\rm vis}(r)$ and $v_{\rm dm}(r)$, for different 
ratios between the Yukawa radius $r_m$ and the disc lengthes $r_d=r_1,r_2$. 
Notice the apparent similarities between the (lower) curves  for $v_{\rm dm}(r)$
in Fig. \ref{Comparison}   and the ones in Fig. \ref{nfw_2} obtained 
for the IT and Burkert halos in the case of the normal gravity. 
\begin{figure}
    \includegraphics[scale=.37]{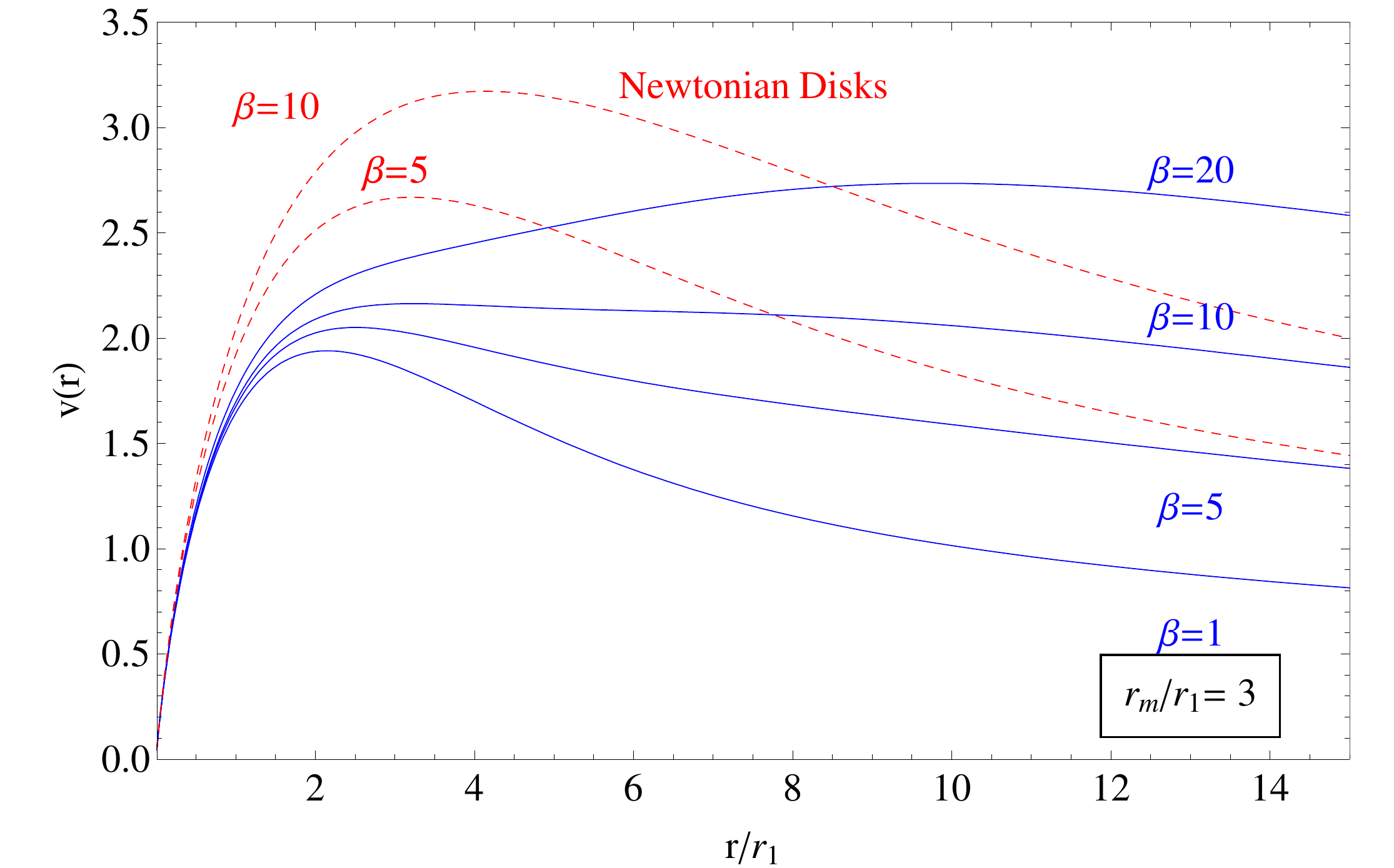}
        \includegraphics[scale=.37]{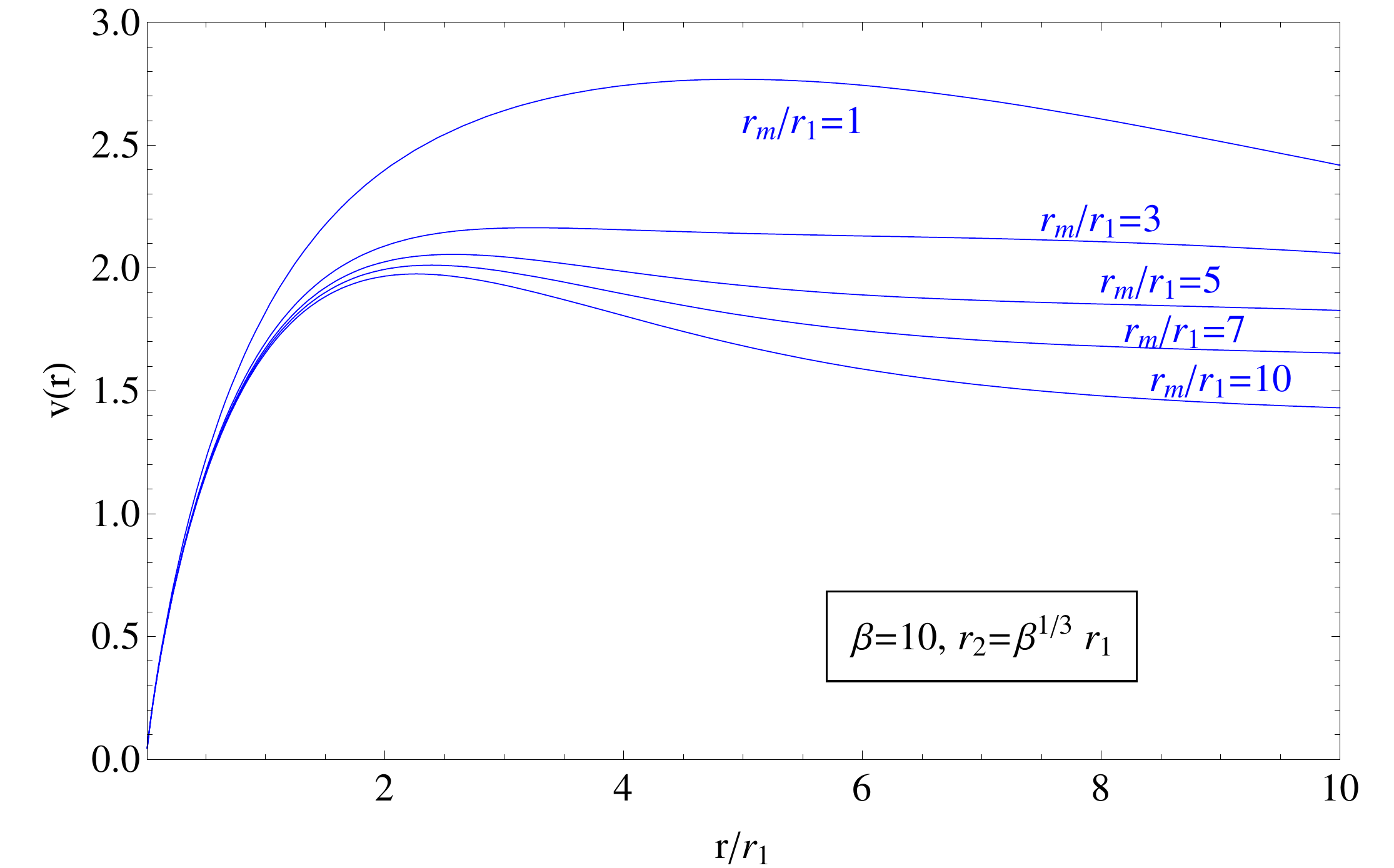}
    \caption{Rotational curves  for different values of   $\beta = M_2/M_1 $ 
    with $\nu = r_m/r_1$ fixed (upper panel) and for different values of $\nu$ 
    with $\beta$ fixed (lower panel). 
  The dotted curves refer to the velocity curves originated by the superimposed 
  dark and visible discs in the case  of unmodified  Newtonian potential.
         }
             \label{p2}
\end{figure}

It is convenient to introduce the rescaling factors 
between the characteristics of the ordinary and dark discs: 
\begin{equation}\label{alpha}
 r_2=\alpha \, r_1 \, , \qquad M_2 = \beta \, M_1 \; .
\end{equation}
%
The parameters $\alpha$ and $\beta$ can differ from galaxy to galaxy, 
while the gravitational radius $r_m$ must be  the same for all galaxies. 
In fact, the values of $\alpha$ and $\beta$ determine the weights by which 
the curves for $v_{\rm vis}(r)$ and $v_{\rm dm}(r)$ shown on Fig. \ref{Comparison} 
for different $r_m/r_{1,2}$ should be superimposed quadratically 
for obtaining the rotational velocity $v^2(r)$ (\ref{vis-dm}). 
Hence, the shape of the rotational curve produced by  the modified gravitational potential  
(\ref{bigr2}) for a given galaxy should be controlled 
by the three ratios $\beta =M_2/M_1$, $\alpha = r_2/r_1$ and $\nu=r_m/r_1$.   

\begin{figure}[t]
\includegraphics[scale=.30]{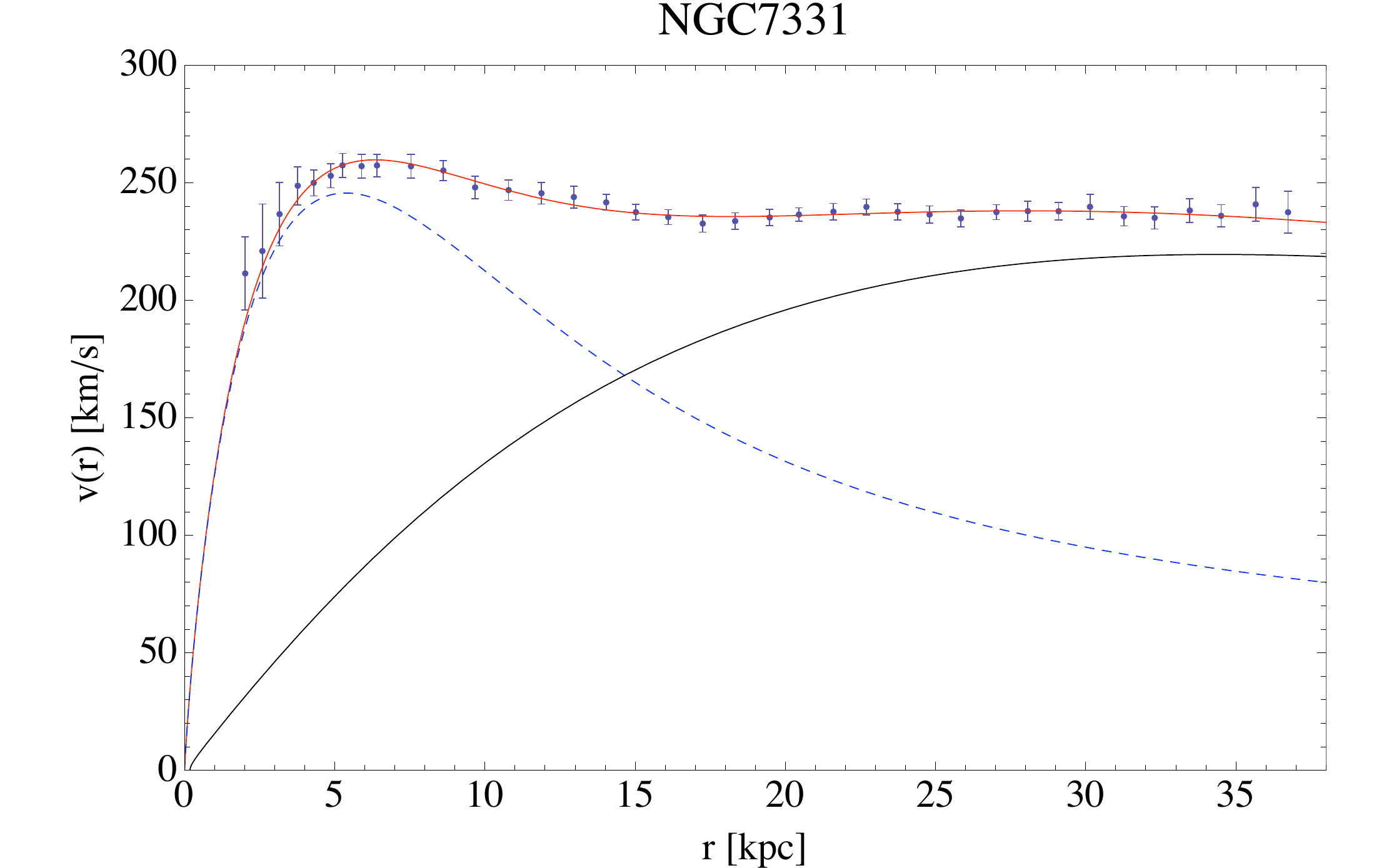} \\
    \includegraphics[scale=.30]{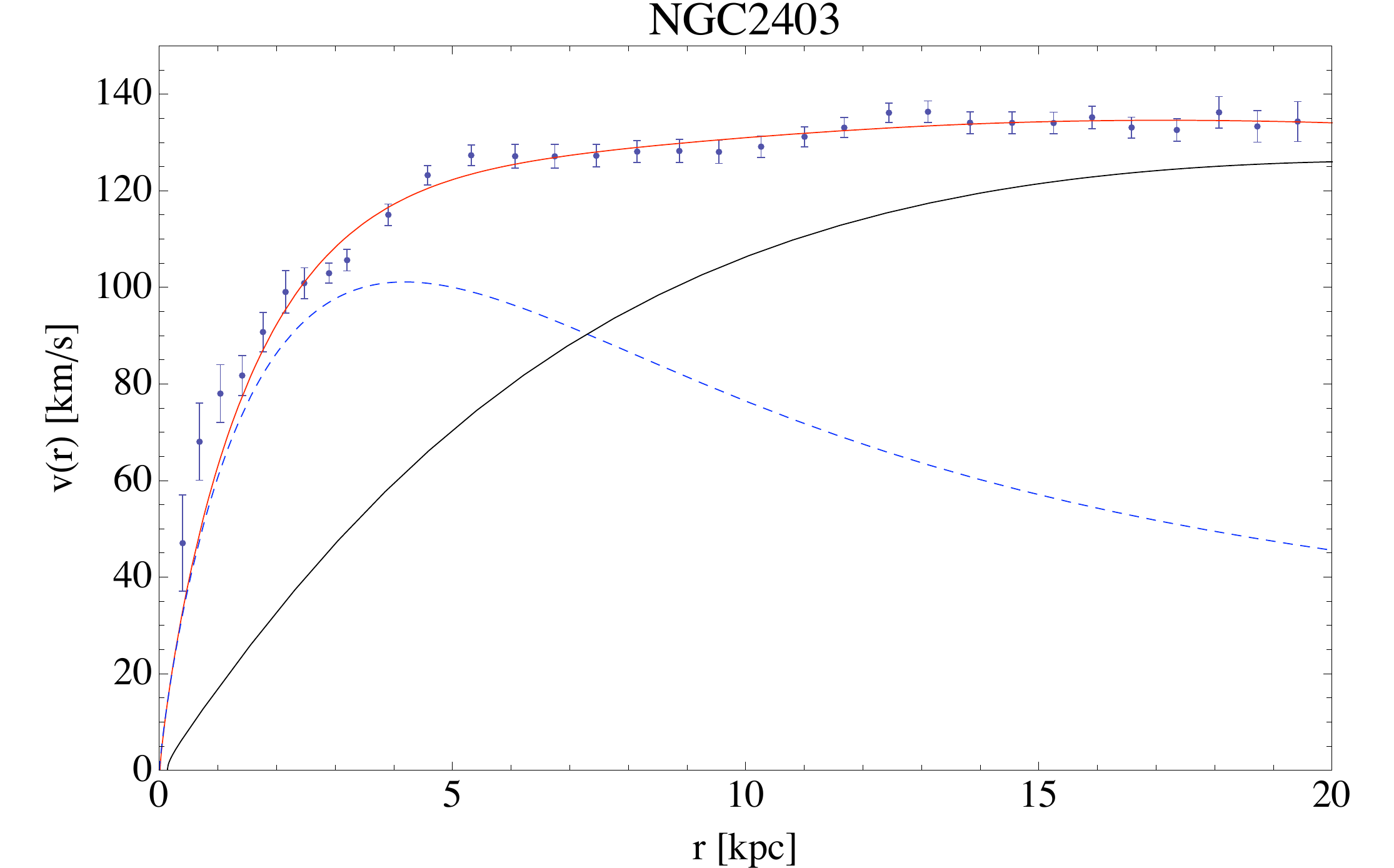} \\
     \includegraphics[scale=.30]{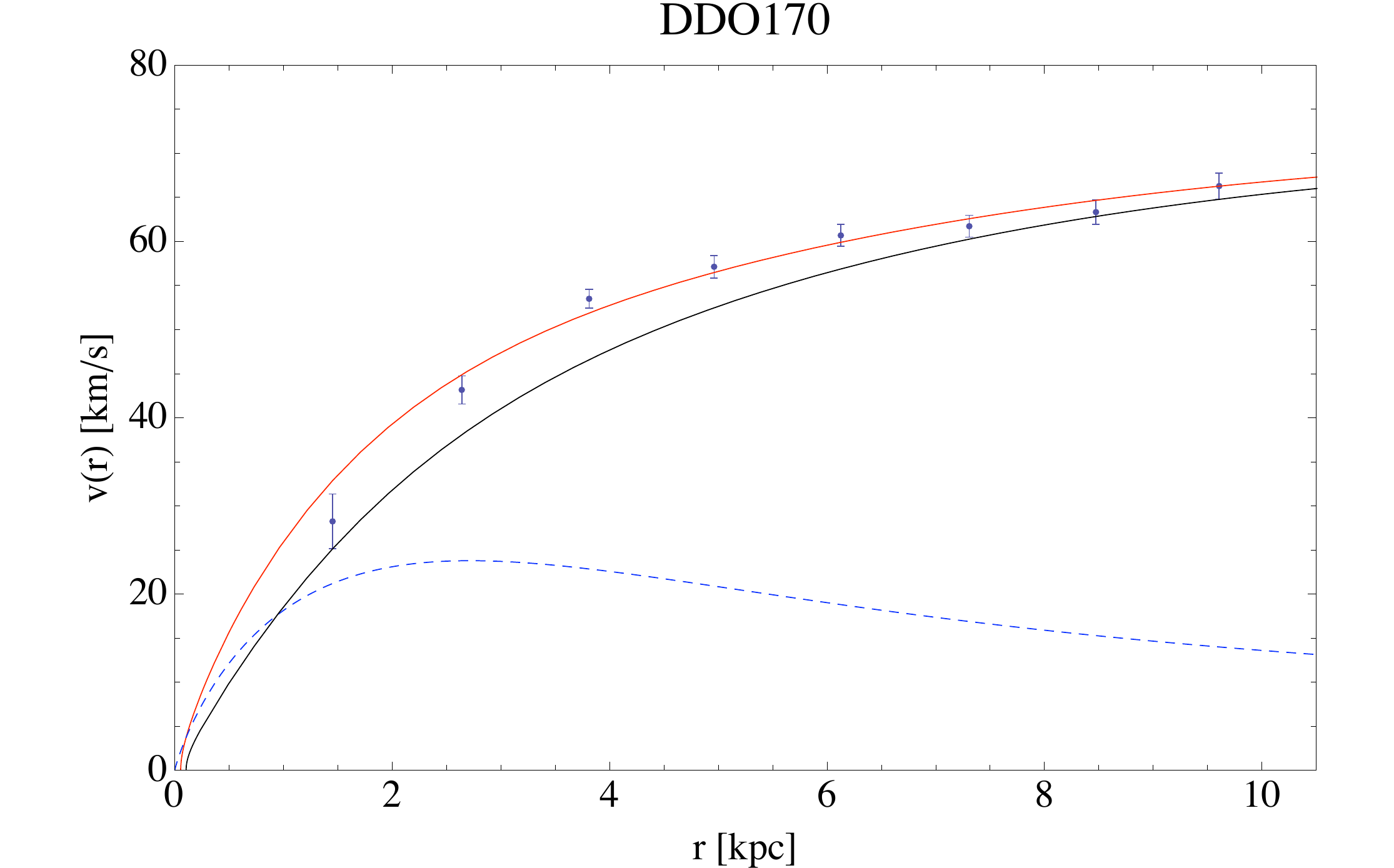} \\
      \includegraphics[scale=.30]{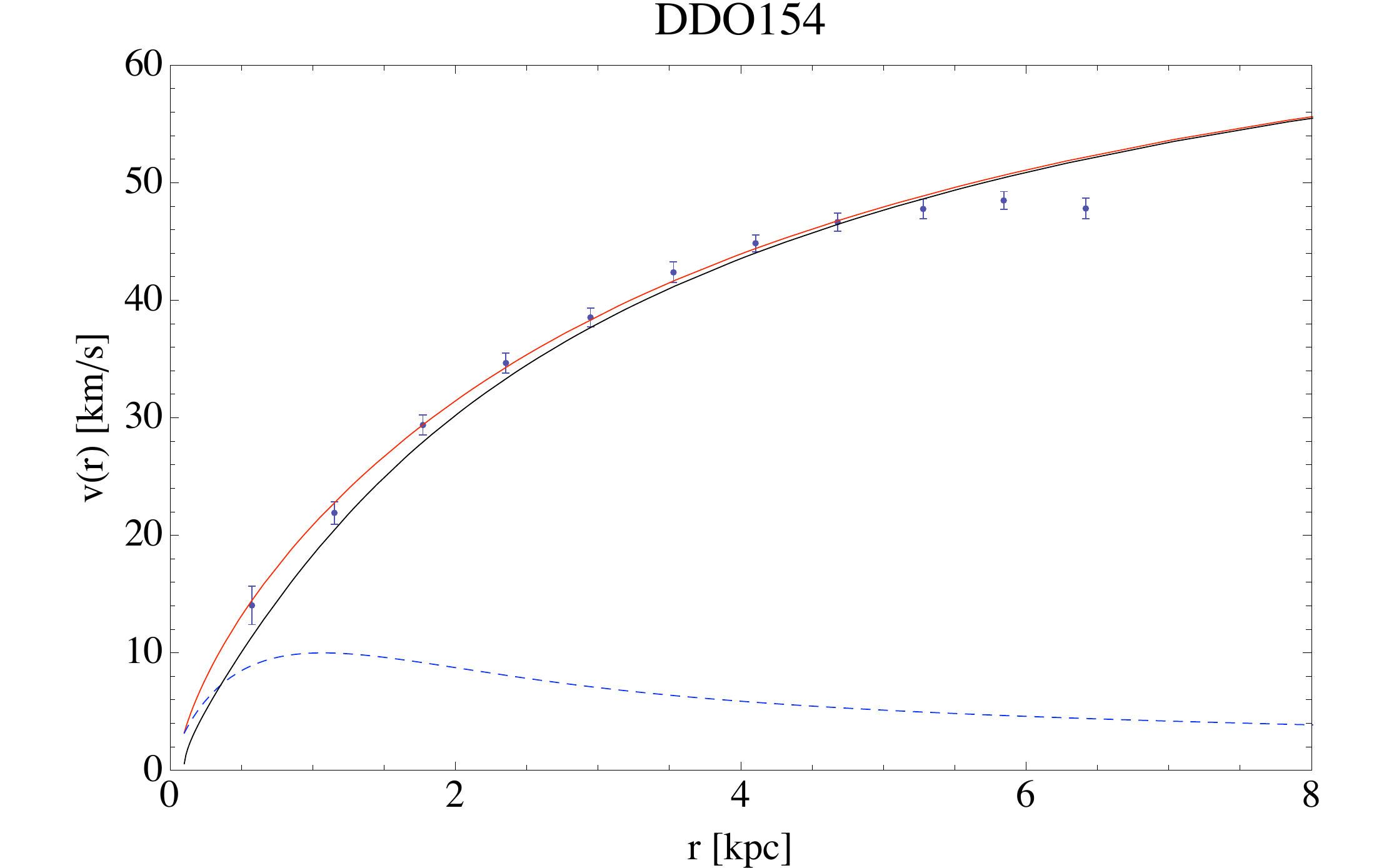}
    \caption{Fit of the rotational curves (solid) for different galaxies 
    using the modified potential and a disc-like distribution for dark matter. 
  In all cases  the Yukawa  radius is  fixed as $r_m=10$ kpc. 
    The  individual contributions  $v_{\rm dm}$ from mirror matter 
and $v_{\rm vis}$   from luminous matter (dotted) are also shown. 
The fit parameters are reported in Table \ref{tab}. }
     \label{NGC2403}
\end{figure}

Rotational curves predicted for different choices of the relevant parameters 
are presented in  Figs. \ref{p2}. 
In particular,  the upper panel of Fig. \ref{p2}  shows how the 
shapes of rotational curves depend on $\beta $ for a fixed $\nu=3$, 
i.e. when the graviton radius $r_m$ is taken equal to three times the disc length scale, 
which is about the optical radius of the galaxy.  For $r_m \sim 10$ kpc, this would 
correspond to the case of the reasonably large galaxies like the Milky Way. 
On the other hand, for large galaxies the ratio $\beta=M_2/M_1$  
is expected to be close to the cosmological ratio between the dark matter and baryonic densities  
while in our model $\beta_{\rm cosm} = \rho_2/\rho_1 = 10$. 
As for the smaller galaxies,  they are expected to be more dark matter dominated, 
with $\beta> 10$, as we discuss below. 
The lower panel of Fig. \ref{p2} shows the dependence on the ratio $\nu=r_m/r_1$, 
for a characteristic value $\beta=10$. 
 In both Figs.  \ref{p2} we assumed that the visible and dark disc density profiles 
 are `zoomed' as $\alpha = \beta^{1/3}$. 
Let us to stress that when gravity is not modified,  it is impossible to reproduce the
observed rotational curves with both dark and visible discs having the similar density 
profiles (\ref{distr}): the Keplerian tail is always there  --  see the dotted curves in Fig. \ref{p2}.

 \begin{table*}[t]
\begin{tabular}{l||c|c|c|c|}
\, & NGC7331 & NGC2403  & DDO170 & DDO154 \\
\hline \hline 
$r_1$ & 4.5 kpc  & 2.0 kpc & 1.1 kpc  & 0.50 kpc  \\
\hline
$M_L$ & 54  $M_{9}$ & 7.9  $M_9$ & 0.18 $M_9$   & 0.05 $M_9$  \\
\hline \hline
$M_1$ & 98  $M_{9}$ & 12.6  $M_9$ & 0.4  $M_9$   & 0.03 $M_9$  \\
$\gamma=M_1/M_L$ & 1.8 & 1.6   & 2.2  & 0.6   \\
\hline 
$M_2$ & 103  $M_{10}$ & 26  $M_{10}$ & 7.6 $M_{10}$   & 6.0 $M_{10}$  \\
$\beta=M_2/M_1$ & 10.5   & 20.6  & 190   & 2000  \\
\hline 
$r_2$ & 9.0 kpc & 3.2 kpc & 0.87 kpc  & 0.66 kpc     \\
$\alpha=r_2/r_1$ & 2.0  & 1.6 & 0.8  & 1.3    \\
\hline \hline
 $\chi^2_{\rm d.o.f.}$ & 0.8 & 1.3 & 1.5   & 0.6   \\
\hline
\hline
\end{tabular}
\caption{The fitting results for different galaxies. Their optical disc length scales $r_1$ and  
the ``luminosity" related masses $M_L=(L/L_\odot)\, M_\odot$ are taken from 
\cite{begeman},  in units defined as   $M_9 = 10^9 \, M_\odot$, etc.; 
$M_1,M_2$ and $r_2$ are the best fit parameters 
(the contribution of gas is included in $M_1$). 
For  NGC7331 also the bulge contribution was taken into account 
as in ref. \cite{begeman}.  
}
 \label{tab}
\end{table*}

Using the modified potential (\ref{bigr2}) and assuming a disc-like density profiles 
for both dark and luminous matter, the rotational curves of different 
(standard, intermediate and dwarf) galaxies can be well reproduced. 
As an example, we fit   the rotational curves of a number of representative  disk galaxies  
(see  Figs. \ref{NGC2403}). 
Among the four parameters, $M_1$,  $r_1$ and $M_2$,  $r_2$, 
the last two regard the dark matter distribution and 
are taken as free parameters for fitting the rotational curve of each galaxy. 
The value of the visible disc length $r_1$ for each galaxy 
can be deduced by  astrophysical measurements,
by extrapolating  the optical shape of the galaxy via exponential profile. 
In other words, the radial distribution of baryonic mass is assumed to be 
given by the mean radial distribution of light;  which is to say,  
the mass-to-light ratio  $\gamma = M_1/M_L$ is taken to be constant 
for each particular galaxy, 
where the value of the luminous mass  $M_L$ 
is inferred from a total luminosity of the galaxy by 
taking   the average mass-to-luminosity  ratio for its stellar population 
equal to that of the sun. 
Therefore, the factor $\gamma$ that takes 
into account the possible variation of the mass-to-light  ratio between 
different galaxies is  taken as a fit parameter, as in ref. \cite{begeman}. 
It is natural to expect that for realistic fits its value should not  strongly deviate from 1. 

In particular,  we fit the  data relative to a large (NGC 7331) and 
an intermediate (NGC 2403) mass spiral galaxies,  as well as 
for two dwarfs (DDO154 and DDO170) which presumably 
are dark matter dominated.  The length scales of these galaxies 
are inferred from their luminosity shape \cite{begeman}. 
We find that all the above galaxies can be nicely fitted by  the three parameters 
$\alpha = r_2/r_1$,  $\beta = M_2/M_1$  and $\gamma = M_1/M_L$, 
with $\chi^2_{\rm dof} \simeq 1$.  
(Let us remark that  for the  standard dark matter halos with the IT or NFW 
profiles also 3-parameter fit is needed, with  $\rho_s$, $r_s$ and 
$\gamma=M_1/M_L$ being the relevant parameters.)
The gravitational radius is {\it a priori} fixed to $r_m=10$ kpc.\footnote{
This choice is essentially based on our intuitive guess. For a complete study, 
a large sample of different galaxies should be analyzed  
by taking $r_m$ as a free parameter that is {\it unique} for all galaxies. 
Such an analysis may suggest that the best fit value for $r_m$ 
is  different from 10 kpc. However, we do not expect the difference more 
than a factor 1.5 or so, i.e. $r_m = 6 \div 15$ kpc.   
}

The fit details  are  shown in Fig.  \ref{NGC2403} and Table \ref{tab}. 
We see that  the smaller galaxies are more dominated by dark matter,   
which can have an intriguing link with the empiric Tully-Fisher relations.  
Namely,  for a large galaxy with $M_1 \sim 10^{11}~ M_\odot$ we have $\beta \sim 10$, 
but  for a galaxy with $M_1 \sim 10^7~M_\odot$ we get $\beta \sim 10^3$. 
This seems quite  natural in the context of our model with 
the Yukawa length  $r_m \sim 10$ kpc.  At the scales smaller than $r_m$ 
 the dark and ordinary components become reciprocally ``blind" and  
 so  the smaller is the dark matter clump, less it should be capable  
to  capture and hold gravitationally the ordinary baryons. 
This feature can be translated to the  following paradigm: 
 big galaxies are expected have the ratio  $\beta=M_2/M_1$ 
nearly approaching the cosmological proportion between 
between dark and visible matter as in cosmology: 
$\beta_{\rm cosm}=\rho_2/\rho_1 \simeq  10$ in our case.   
As for small galaxies, the value of $\beta$ can be substantially larger: 
the smaller is the dark matter disc, the less amount of  the ordinary baryons 
it can capture and hold against the galaxy scattering and merging processes  
and thus should be more dark matter dominated.  

For scales $r >r_m$ gravity is essentially Newtonian, with a constant $G_{\rm N}/2$.  
Therefore, for  normal galaxies one can expect the evidence 
of quasi-Keplerian fall-off at the distances $r > 5 r_m$. 
In principle, such a feature may be traced in large galaxies where the 
velocities can be measured beyond  $50$ kpc or so.  

Certainly, our scenario of  gravity modification can be applied also 
when dark matter is cold and collisionless  (CDM-like)  and forms extended halos 
instead of the discs. In this case, for fitting the rotational curves,
the gravitational radius  presumably will be somewhat less than 
10 kpc, and the rotational curves can be reproduced in the Newton-like 
large distance limit of the potential (\ref{bigr2})  with a constant $G_{\rm N}/2$. 
Therefore, the value of $\beta = M_2/M_1$ is required to be twice 
bigger than in the canonical  CDM interacting with Newton constant  $G_{\rm N}$.
The gravity  screening between ordinary and dark components at distances $r<r_m$ 
would help to avoid the cusp problem: 
even if the dark matter density distribution has a central singularity, 
it would not affect the rotational curves at small distances.

Notice that, taking $r_m \sim 10$ kpc, for  the galaxy 
clusters we recover the standard picture if  the dark-to-baryon mass 
ratio in the cluster is  $\beta = 10$ or so.  
Indeed,  the relevant length scales in a cluster are bigger than $r_m$ 
and  the potential (\ref{bigr2})  reduces to the   Newtonian
potential but with a half-strength $G_{\rm N}/2$. 
Therefore, the velocity dispersions observed in clusters require 
$M_{2} \simeq 10 M_1 $, instead of $M_{2} \simeq 5 M_1 $ 
which works when the Newton constant is $G_{\rm N}$.

\section{Discussion and outlook}

Mirror matter is a dark matter candidate that can have interesting applications 
for many cosmological problems \cite{alice,fractions}.  
However, being collisional and dissipative,  it has difficulties with the formation of 
extended halos, and thus with the explanation of the galactic  rotational velocities.  
Typically, for fitting  the rotational curves without modifying the Newton
potential,   dark matter profile with a constant inner core is required, 
as e.g.  in refs. \cite{begeman,burkert}. 

 In this paper we have shown  that  the flat rotational curves can be nicely 
 explained  by  the bi-metric  modification of gravity suggested in ref. \cite{trigravity}  
 where the hidden mirror sector is equipped by its own mirror gravity. 
In other words, if the Universe is made of two separate gauge sectors, one visible and
the other mirror, each of them can have its own gravity. The two metrics can 
interact leading to a large distance Yukawa-type modification of the gravitational force. 
At  small distances, where the gravitational interaction between the different 
sectors is effectively shut off, the behavior is practically Newtonian. 

Though the structure formation process is very complicated,  some
simple conclusions can be drawn.  Dark matter  is more 
abundant and it will start collapsing before the ordinary matter 
forming gravitational wells for capturing the latter. 
In difference from the CDM,  the collisional MDM is is tended to 
undergo  a dissipative collapse creating structures as galaxies,  
similar to the visible ones, with similar mass distribution. 
Hence, it is natural to expect that similarly to the visible matter, mirror matter 
forms galactic discs.  
Since the gravitational force between sectors 1 and 2 is effective 
at distances  $r \gtrsim  r_m \sim 10$ kpc,  after an initial stage of mutual 
interaction,  as the  collapse proceeds, 
the inner structures of  dark and ordinary matter are formed with a  
weaker correlation. 
On the contrary, one can argue that for the outer region of 
galaxies,  the interaction between ordinary and dark matter is important for 
angular momentum exchange  giving rise to a common galactic plane. 
  
It is important, that a good dark matter candidate together with a plausible 
gravity modification should reproduce the observed  rotational curves 
for the galaxies of different size and type.  
We have shown that  the rotational curves can be explained if  
the dark and visible matter components both have  similar mass distributions 
in discs, with exponential density profiles 
 $\sigma_{1,2}(r)\propto e^{-r/r_{1,2}}$, where $r_1,r_2$ respectively 
 are the length scales of the visible and dark discs.  
 Let us remark, however, that our proposal   of  gravity modification  
 would work  even if dark matter behaves as the CDM,  
 avoiding the tension between the observed rotational curves and the cusped halos  
 predicted by  N-body numerical simulations. 
  The idea is that  dark matter does not 
 gravitate with normal matter at small distances and the dark matter cusp 
 would not affect the shape of velocity curves. 

Observations also show that in a  galaxy the  dark-to-visible matter ratio   
fractions sharply  increases with decreasing luminosity (see Table \ref{tab}). 
This property is natural in our model, where the graviton Compton length 
$r_m$ is of order $10$ kpc and at smaller scales the dark and ordinary 
components become rather ``blind" to each other. Therefore, the 
smaller is the dark matter disc, the less amount of  the ordinary baryons 
it can capture and thus should be more dark matter dominated. 
For scales $r \gg r_m$ gravity is essentially Newtonian and 
in very large galaxies with an optical radius greater than $50$ kpc or so 
the rotational curves should show the quasi-Keplerian fall-off. 
This can be  interpreted in standard 
paradigm as a fact that the bigger galaxies have less fraction of dark matter, 
approaching the cosmological proportion  between $\rho_D$ and $\rho_B$ 
(about 10 in our case), 
while the small galaxies should be more dominated by dark matter.

\begin{figure}[t]
  \hskip -.3cm
    \includegraphics[scale=.38]{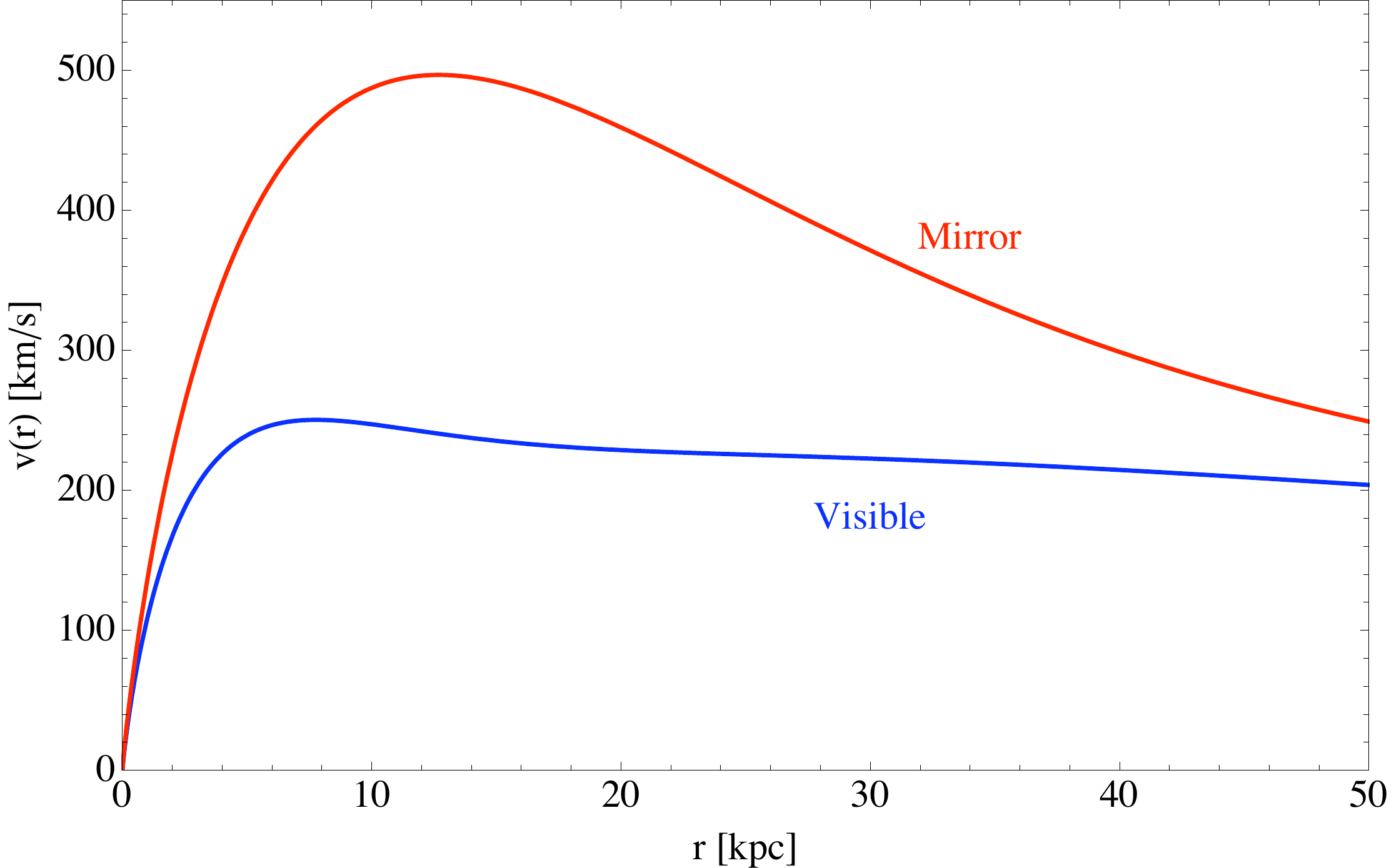}
     \caption{Velocity curves  for ordinary and mirror particles in a typical galaxy 
     similar to the Milky Way, with   $M_1=5 \times 10^{11}\, M_\odot$ and $r_1=3$ kpc. 
     For a dark disc, we take   
     $M_2=5 \times 10^{12}\, M_\odot$ and $r_2=7$ kpc. 
     The enhancement of  the dark matter velocity is important for its direct detection.}
	 \label{p3}
\end{figure} 

The following remark is in order. As far as the galaxies and clusters 
are regarded,  the ``Newtonian" character of our  gravity 
at large distances, with  $G(r> r_m) = G_{\rm N}/2$  but $\beta \sim 10$,  
will have the similar effect as  the  extended dark matter halos also concerning the 
gravitational lensing as far as it essentially imitates  
the similar shape of the gravitational potential. As for the microlensing events 
by Machos and other compact objects in the Milky Way and its neighborhoods,   
were the relevant distances are of the order of the stellar size at which 
the dark matter remains gravitationally ``invisible", 
the corresponding dynamics should be governed exclusively 
by the standard gravity  with the canonical  Newton constant  $G_{\rm N}$.  

Intriguingly, recent astrophysical observations 
indicate a peculiar behavior of  dark matter   in galaxy cluster collisions. 
In the Bullet cluster~\cite{bullet}, the dark matter component of the cluster 
shows a collisionless  behavior. Instead, in the Abell 520 cluster~\cite{abell} 
there is an evidence for an important contribution from self-interaction. 
In mirror dark matter models those observations are not surprising~\cite{silag}.

At large cosmological distance  both sectors are mutually interacting  
with an effective Newton constant $G_{\rm N}/2$.
One can argue then that the observed Hubble constant 
requires that the total energy density of the Universe is
twice bigger than in the standard cosmology in which the Newton
constant at cosmological distances remains the canonical $G_{\rm N}$.
In this scenario, instead of $\rho\approx \rho_{cr}=3H_0^2/8\pi G_{\rm N}$ of the 
standard FRW cosmology, we expect 
 $\rho_{our} \approx 2\rho_{\rm cr} =3H_0^2/4\pi G_{\rm N}$.

The mirror matter model with modified potential does not exclude
the possibility of a direct (non-gravitational) interaction
between  normal and dark matter components.  
For example, the  photon - mirror photon kinetic mixing 
$\frac{\epsilon}{2} F^{\mu\nu}_1 F_{2\mu\nu}$ \cite{photons} 
that effectively renders mirror particles `millicharged' with respect to ordinary 
electromagnetic interactions. In this way,  mirror nuclei can scatter off the 
ordinary nuclei in the experimental set up for the dark matter direct detection 
and thus leave their trace on the recoil energy spectrum of the latter. 
This process 
might  be suitable for explaining the results  of the DAMA/Libra  experiment 
\cite{bernabei08}, if  $\epsilon \sim 10^{-9}$ and the mirror sector is dominated 
by the oxygen or other mirror elements with comparable mass  \cite{footd}. 
However, such a large value of $\epsilon$ seems to be in tension 
with the cosmological limits on the mirror particle electric charges 
\cite{Berezhiani:2008gi} while the large oxygen fraction  is also questionable.  

It is worth to point out that in our  model the observed shape 
of the velocities requires at least twice the amount
of dark matter in our galaxy than the CDM  scenario.  
In addition, if it is distributed in a disc, then the local density of dark matter 
should be much larger than in the case of extended quasi-spherical halos. 
On the other hand,  gravity is not universal in our model, and thus ordinary and mirror 
objects in the galaxy should have different accelerations. 
Namely, the typical velocities of  dark matter particles in the Milky Way at the distances 
of 10 kpc in our scenario approach 500 km/s, while the 
velocities of the ordinary matter (and sun itself) are around 200 km/s 
(see Fig. \ref{p3}).  
Both of these features are of interest  for dark matter searches, 
since they enhance the chance of a direct detection in experiments like DAMA 
\cite{bernabei08}. 
Namely, if the MDM has a local density much larger than the one 
inferred from the CDM halo profile, this 
 would make the smaller values $\epsilon \ll 10^{-9}$ effective for dark 
matter detection. 
On the other hand, 
if the typical velocities of dark matter particles are more than twice 
larger than what is assumed in the  case of CDM  halos, 
then for producing  a signal with a recoil 
energies in the range 2-6 keV, the "bigravitating" dark matter particles may be lighter 
than normally gravitating dark matter particles for which 
the   typical velocities around 200-300 km/s are inferred. 
In particular,  instead of mirror oxygen, the element that  becomes operative 
is mirror helium, which is expected to be a dominant component in mirror 
sector \cite{BCV}. 

Our scenario can have also rather interesting implications 
for neutrino-mirror neutrino (active-sterile) oscillations
\cite{neutrinos}, making more attractive the case of an exact mirror parity 
where the  ordinary and mirror neutrinos are degenerate in mass and maximally mixed.
Their oscillations inside the galaxy should suffer an MSW-like
suppression because of different gravitational potentials felt by active and
sterile neutrino species. Far outside the galaxy where the gravity becomes
universal oscillations instead can proceed in full strength.


\subsection*{Acknowledgments}
\noindent
We thank F. Nesti and P. Salucci for useful discussions and comments.
This work is partially supported 
by the European FP6 Network "UniverseNet" 
MRTN-CT-2006-035863 and in part by the MIUR biennal 
grant for the Research Projects of the National Interest PRIN 08 
on "Astroparticle Physics". 


\end{document}